\documentclass[nofootinbib]{revtex4-1}%

\usepackage{amssymb}
\usepackage{amsfonts}
\usepackage{amsmath}
\usepackage{graphicx}

\renewcommand{\d}{\partial}

\renewcommand{\geq}{\,{\geqslant}\,}
\renewcommand{\leq}{\,{\leqslant}\,}

\newcommand{\binner}[2]{%
  {\langle}\kern-4.15pt{\langle}#1{,}\,#2{\rangle}\kern-4.15pt{\rangle}}

\newcommand{\half}{\mathchoice{%
    \ffrac{1}{2}}{\frac{1}{2}}{\frac{1}{2}}{\frac{1}{2}}}

\newcommand{\ffrac}[2]{\raisebox{.5pt}%
  {\footnotesize$\displaystyle\frac{#1}{#2}$}\kern1pt}

\DeclareFontFamily{OT1}{rsfs}{} \DeclareFontShape{OT1}{rsfs}{m}{n}{
<-7> rsfs5 <7-10> rsfs7 <10-> rsfs10}{}
\DeclareMathAlphabet{\mycal}{OT1}{rsfs}{m}{n}
\def\scri{{\mycal I}}%
\def\scrip{\scri^{+}}%

\begin{document}

\title{The flat limit of three dimensional asymptotically anti-de
  Sitter spacetimes}



\author{Glenn Barnich} \email{gbarnich@ulb.ac.be} \thanks{Research
  Director of the Fund for Scientific Research Belgium}
\affiliation{Physique Th\'eorique et Math\'ematique, Universit\'e
  Libre de Bruxelles and International Solvay Institutes, Campus
  Plaine C.P. 231, B-1050 Bruxelles, Belgium}

\author{Andr\'es Gomberoff}
\email{agomberoff@unab.cl}
\affiliation{Universidad Andres Bello, Departamento de Ciencias F\'{\i}sicas,\\
 Av. Rep\'{u}blica 252, Santiago,Chile. }

\author{Hern\'an A. Gonz\'alez}
\email{hdgonzal@uc.cl}
\affiliation{Departamento de F\'{\i}sica, P. Universidad Cat\'olica de
Chile, Casilla 306, Santiago 22,Chile}



\begin{abstract}
  In order to get a better understanding of holographic properties of
  gravitational theories with a vanishing cosmological constant, we
  analyze in detail the relation between asymptotically anti-de Sitter
  and asymptotically flat spacetimes in three dimensions. This
  relation is somewhat subtle because the limit of vanishing
  cosmological constant cannot be naively taken in standard
  Fefferman-Graham coordinates. After reformulating the standard
  anti-de Sitter results in Robinson-Trautman coordinates, a suitably
  modified Penrose limit is shown to connect both asymptotic regimes.
\end{abstract}


\maketitle

\section{Introduction}

Even though quantum properties are still not completely understood
\cite{Maloney:2007ud,Witten:2007kt}, on the classical and
semi-classical level asymptotically anti-de Sitter spacetimes in three
dimensions constitute an extremely rich and well-studied framework:

\begin{enumerate}
\item As an early precursor to the AdS/CFT correspondence
  \cite{Maldacena:1997re,Aharony:1999ti}, their symmetry algebra has
  been shown to consist of two commuting copies of centrally
  non-extended Virasoro algebras with a central extension arising in
  the Dirac bracket algebra of the canoncial generators
  \cite{Brown:1986nw}. The value of the central charges
  $c^\pm=\frac{3l}{2G}$ has been used to argue for a microscopic
  derivation of the Bekenstein-Hawking entropy of the black holes
  \cite{Strominger:1998eq}, independently of the details of the
  underlying theory \cite{Martinec:1998wm,Carlip:1998qw}.

\item The general solution to the equations of motion is known in
closed form \cite{1999AIPC..484..147B,Skenderis:1999nb} in
Fefferman-Graham coordinates, it includes
spinning black holes \cite{Banados:1992wn}, and all other solutions
can be obtained from this 2 parameter family through suitable
coordinate transformations \cite{Rooman:2000ei}.  

\item Additional control on holographic properties comes from the
  Chern-Simons formulation of three dimensional anti-de Sitter gravity
  \cite{Achucarro:1986vz,Witten:1988hc} and the relation of
  Chern-Simons theories with conformal field theories on their
  boundary \cite{Witten:1988hf,Moore:1989yh,Elitzur:1989nr}. In
  particular, the asymptotic dynamics has been understood from this
  point of view in \cite{Coussaert:1995zp,PhysRevD.52.5816} (see also
  \cite{Balachandran:1991dw}).
\end{enumerate}

In view of these results, a valid strategy to get additional insight
into holographic properties of gravitational theories with vanishing
cosmological constant
\cite{witten:98xx,Susskind:1998vk,Polchinski:1999ry,deBoer:2003vf%
  ,Solodukhin:2004gs} is to study in more detail asymptotically flat
spacetimes in three dimensions and their relation to the
asymptotically anti-de Sitter case.

The symmetry algebra $\mathfrak{bms}_3$ of asymptotically flat three
dimensional spacetimes \cite{Ashtekar:1996cd} involves both a
supertranslation and a superrotation sub-algebra, the latter being
given by a centrally non-extended Virasoro algebra.  The Dirac bracket
algebra of the surface charges has a central charge with value
$c=\frac{3}{G}$ between the superrotation and supertranslation
generators \cite{Barnich:2006avcorr}, which is related to the anti-de
Sitter algebra through a suitable redefinition of the generators
followed by taking the cosmological constant to zero. Furthermore,
using the three dimensional analog of the four-dimensional
Bondi-Metzner-Sachs (BMS) gauge \cite{Bondi:1962px,Sachs:1962wk}, the
general solution to the equations of motion can also be found in
closed form \cite{Barnich:2010eb}. It involves two arbitrary functions
of one variable, exactly as in the anti-de Sitter case. Finally, from
the Chern-Simons point of view, some aspects of the boundary dynamics
of flat space gravity have been discussed in \cite{Salomonson:1989fw}.

In this paper we study in detail the suitably modified Penrose limit
that connects the general solution, symmetries and surface charges of
asymptotically $AdS_3$ and Minkowski spacetimes in three dimensions.

The plan of the paper is as follows. In the next section, we briefly
recall standard results on asymptotic symmetries, solutions and
charges in the anti-de Sitter case in Feffermann-Graham coordinates
and show that the limit $l\to\infty$ cannot be performed in a
straightforward way. In section III, we re-derive the anti-de Sitter
results in the BMS gauge. The procedure for taking the limit is
explained in detail in section IV. In section V, we briefly discuss the simplest
geometries when only the zero modes are excited. 

The flat limit from the point of view of the Chern-Simons formulation,
including a non-relativistic $\mathfrak{bms}_3$ invariant Liouville theory,
will be discussed elsewhere. More generally, the representation theory
for $\mathfrak{bms}_3$ needs to be studied in more detail. An
investigation using the fact that $\mathfrak{bms}_3$ is isomorphic to
$\mathfrak{gca}_2$, a non-relativistic contraction of the Virasoro
algebra, can be found in
\cite{Bagchi:2009pe,PhysRevLett.105.171601,Bagchi:2012cy}. 

\section{Symmetries, solutions and charges in the Fefferman-Graham
  gauge}

An asymptotically $AdS_3$ metric in the spirit of Fefferman-Graham
consists in a metric ansatz
\begin{equation}
ds^2=\frac{l^2}{\rho^2}d\rho^2+g_{AB}(\rho,x)
dx^Adx^B,\label{eq:7}
\end{equation}
where the negative cosmological constant is $\Lambda=-1/l^2$.  The
Einstein equations of motion then imply in particular that
\begin{equation}
  \label{eq:A9}
  g_{AB}=\rho^2\bar\gamma_{AB}+O(1),
\end{equation}
and we take for simplicity $\bar\gamma_{AB}=\eta_{AB}={\rm
  diag}(-1,1)$, the flat metric on the cylinder $x^A=(\frac{t}{l},\phi)$ in what
follows. It will also be useful to use light-cone coordinates
$x^\pm=\frac{t}{l}\pm\phi$.

The infinitesimal transformations leaving the form \eqref{eq:7} and
\eqref{eq:A9} of the metric invariant are generated by
\begin{eqnarray}
\left\{\begin{array}{l}\xi^\rho=-\half\psi \rho, \\ \xi^A  =  Y^A +
    I^A,\quad 
 I^A=-\frac{l^2}{2}\d_B
  \psi\int_\rho^\infty \frac{d\rho^\prime}{\rho^\prime}
  g^{AB},
\end{array}\right.\label{eq:FGvect}
\end{eqnarray}
where $Y^A$ is a conformal Killing vector of $
\bar\gamma_{AB}dx^Adx^B=-dx^+dx^-$, while $\psi=\bar D_A Y^A$ is
the conformal factor.  These spacetime vectors form a representation
of the algebra of conformal Killing vectors of
$\bar\gamma_{AB}dx^Adx^B$,
\begin{equation}
  \label{eq:44a}
  [\xi_{Y_1},\xi_{Y_2}]^\mu_M \equiv[\xi_{Y_1},\xi_{Y_2}]^\mu-
\delta^g_{\xi_{Y_1}}\xi^\mu_{Y_2}+
  \delta^g_{\xi_{Y_2}}\xi^\mu_{Y_1}=\xi^\mu_{[Y_1,Y_2]}. 
\end{equation}
On the cylinder, one can expand $Y^\pm(x^\pm)=\sum_{m\in \mathbb{Z}}
Y^\pm_{m} e^{-i m x^\pm}$, with $\bar Y^\pm_m=Y^\pm_{-m}$. For the
generators $l^+_m=\xi^\mu_{e^{imx^+},0}\d_\mu$,
$l^-_m=\xi^\mu_{0,e^{imx^-}}\d_\mu$ one gets
\begin{equation}
  \label{eq:1}
  i[l^\pm_m,l^\pm_n]_M=(m-n)l^\pm_{m+n},\quad  [l^\pm_m,l^\mp_n]_M=0\,.
\end{equation}

When requiring that $\bar\gamma_{AB}=\eta_{AB}$, the general solution
to Einstein's equations is given by
\begin{equation}\label{FG}
  ds^2=\frac{l^2}{\rho^2}d\rho^2-(\rho dx^+-\frac{l^2}{\rho}\Xi_{--}dx^-)
( \rho dx^--\frac{l^2}{\rho}\Xi_{++}dx^+),
\end{equation} 
where $\Xi_{\pm\pm}=\Xi_{\pm\pm}(x^\pm)$. The easiest solutions where
these functions are constants 
\begin{equation}
\Xi^{BTZ}_{\pm\pm}=2G(M\pm \frac{J}{l}),\label{BTZ}
\end{equation}
include both the BTZ black holes for which $M\geq 0$, $|J|\leq Ml$ and
$AdS_3$ which corresponds to $M=-\frac{1}{8G}$, $J=0$.

The action of the transformations generated by $\xi_{Y}$ on solution
space reads
\begin{equation}  -\delta_Y\Xi_{\pm\pm}=Y^\pm\d_\pm
  \Xi_{\pm\pm}+2\d_\pm Y^\pm \Xi_{\pm\pm}-\half \d^3_\pm
  Y^\pm.\label{eq:A15}
\end{equation} 
The conserved surface charges, computed with respect to the $AdS_3$
background\footnote{In \cite{Barnich:2010eb} an overall factor of $l$
  was missed, while the charges were computed with respect to the
  $M=0=J$ BTZ black hole.}, are given by
\begin{equation}
  \label{eq:2}
  Q_Y=\frac{l}{8\pi G}\int_0^{2\pi}d\phi
  \Big[Y^+(\Xi_{++}+\frac{1}{4})+Y^-(\Xi_{--}+\frac{1}{4})\Big].
\end{equation}
The generators are denoted by $L^+_m=Q_{e^{imx^+},0}$,
$L^-_m=Q_{0,e^{imx^-}}$ and for the BTZ black hole we find in
particular
\begin{equation}
\frac{1}{l}(L^{+ BTZ}_m+ L^{- BTZ}_m)=\delta_m^0(M+\frac{1}{8G}),\qquad
L^{+ BTZ}_m- L^{- BTZ}_m=\delta^m_0 J\label{eq:9},
\end{equation}
the non-vanishing value being associated with $\d_\tau=l\d_t$ and with
$\partial_\phi$ respectively. More generally, up to normalization
and a shift of the zero mode, the  charges $L^\pm_{-m}$ are
the coefficients of the Fourier expansion of $\Xi_{\pm\pm}$, 
\begin{equation}
  \label{eq:11a}
  \Xi_{\pm\pm}=-\frac{1}{4} +\sum_m  \frac{4 G}{l} L^\pm_{m} e^{-imx^\pm}.
\end{equation}

According to general results from
\cite{Regge:1974zd,Brown:1986nw,Brown:1986ed,Barnich:2001jy} the Dirac
bracket of the surface charges is taken to be 
\begin{equation}
  \label{eq:3}
  \{Q_{Y_1},Q_{Y_2}\}=\delta_{Y_1} Q_{Y_2}.
\end{equation}
For the generators, one then gets
\begin{equation}
  \label{eq:4}
   i\{L^\pm_m,L^\pm_n\}=(m-n)L^\pm_{m+n}+\frac{c^\pm}{12}m(m^2-1)
\delta^0_{m+n},\quad
   \{L^\pm_m,L^\mp_n\}=0,
\end{equation}
where 
\begin{equation}
  \label{eq:6}
  c^\pm=\frac{3l}{2G}.
\end{equation}
After quantization (with $\hbar =1$), the commutator of the associated
quantum operators in the limit where $c^\pm\gg 1$ is given by
\begin{equation}
  \label{eq:4b}
   [L^\pm_m,L^\pm_n]=(m-n)L^\pm_{m+n}+
\frac{c^\pm}{12}m(m^2-1)\delta^0_{m+n}
   ,\quad  [L^\pm_m,L^\mp_n]=0.
\end{equation}

By defining the generators
\begin{eqnarray}
  \label{generators}
 P_{m}=\frac{1}{l}(L^{+}_{m}+L^{-}_{-m}),  \quad  J_{m}=L^{+}_{m}-L^{-}_{-m},
 \end{eqnarray}
the surface charge algebra becomes
\begin{equation}
\label{eq:qflat}
\begin{gathered}
i\{J_m,J_n\}=(m-n)J_{m+n}+\frac{c^+-c^-}{12}m(m^2-1)\delta^0_{m+n},\\
i\{J_m,P_n\}=(m-n)P_{m+n}+\frac{c^++c^-}{12\ell}m(m^2-1)\delta^0_{m+n},\\
i\{P_m,P_n\}=\frac{1}{l^2}\big((m-n)J_{m+n}+
\frac{c^+-c^-}{12}m(m^2-1)\delta^0_{m+n}\big).
\end{gathered}
\end{equation}
In the purely gravitational case with central charges as in
\eqref{eq:6}, the limit $l\rightarrow\infty$ is well-defined and gives
rise to the centrally extended $\mathfrak{bms}_{3}$ algebra
\cite{Barnich:2011ct},
\begin{equation}
\label{eq:qflatbis}
\begin{gathered}
i\{J_m,J_n\}=(m-n)J_{m+n}+\frac{c_1}{12}m(m^2-1)\delta^0_{m+n},\\
i\{J_m,P_n\}=(m-n)P_{m+n}+\frac{c_2}{12}m(m^2-1)\delta^0_{m+n},\\
i\{P_m,P_n\}=0,
\end{gathered}
\end{equation}
where
\begin{equation}
c_1=0,\quad c_2=\frac{3}{G}\label{eq:8}.
\end{equation}
This agrees with the results found by a direct computation for
asymptotically flat Einstein gravity in three dimensions in
\cite{Barnich:2006avcorr}.

Note, however, that the line element (\ref{FG}) is not well defined in
the limit $l\rightarrow\infty$. This is not entirely surprising since
taking limits in spacetime geometries is quite a subtle issue
\cite{Geroch:1969xv}. In particular, results depend on the coordinates
that are held fixed during the limit.

\begin{figure}[h]
\label{fig} 
\begin{center}
  \includegraphics[width=10cm]{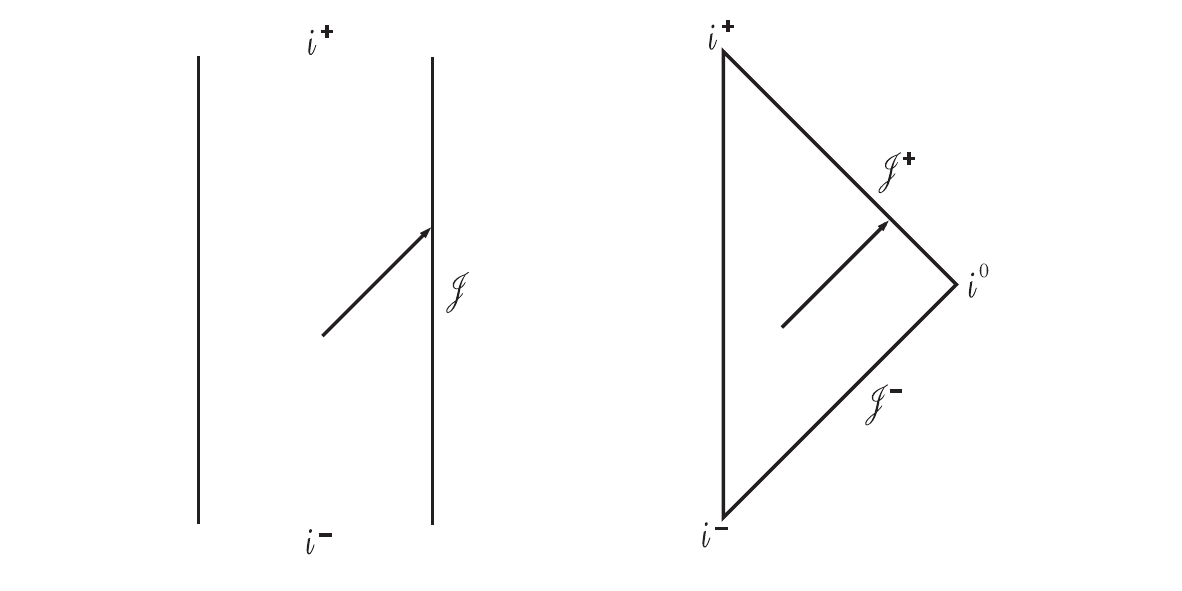}
  \caption{Penrose diagrams of anti-de Sitter and Minkowski
    spacetimes. Arrows in the diagrams represent outgoing null rays. }
   \end{center}
\end{figure}

\section{Results in the BMS gauge}

In order to be able to relate the asymptotic analysis in both cases we
have to take into account that in flat space, the analysis is
performed at null infinity. For definiteness, we concentrate on future
null infinity $\scrip$. Since gauge fixing is independent of the
presence of a cosmological constant, we can choose the three
dimensional analog of the BMS gauge for both asymptotics. This can be
done by making the following metric ansatz in terms of coordinates
$u,r,\phi$,
\begin{equation}
  \label{eq:ojo}
   ds^2=e^{2\beta}\frac{V}{r} du^2-2e^{2\beta}
   dudr+r^2(d\phi-Udu)^2,
\end{equation}
for three arbitrary functions $\beta, V, U$. For instance, defining
the retarded time $u$ through $t=u+l\arctan{\frac{r}{l}}$, the $AdS_3$
metric in global coordinates,
\begin{equation}\label{back}
ds^2=-\Big(1+\frac{r^2}{l^2}\Big)dt^2+
\Big(1+\frac{r^2}{l^2}\Big)^{-1}dr^2+r^2d{\phi}^2,
\end{equation}
is of the form \eqref{eq:ojo} with $\beta=0=U$,
$\frac{V}{r}=-\frac{r^2}{l^2}-1$. Furthermore, the limit $l\to\infty$
can be safely taken and yields Minkowski spacetime, as desired.

Assuming that $\beta=o(1)=U$, the Einstein equations of motion imply
in particular that
\begin{equation}
  \frac{V}{r} = -\frac{r^2}{l^2}+O(1), \quad \beta = O(r^{-1}),\quad
  U= O(r^{-2}). \label{eq:fall1}
 \end{equation}
Infinitesimal transformations that keep the gauge fixed form
 \eqref{eq:ojo} invariant are generated by vector fields $\xi^\mu$
 such that
\begin{equation}
\mathcal{L}_\xi g_{rr}=0=\mathcal{L}_\xi g_{r\phi}= \mathcal{L}_{\xi}
g_{\phi\phi}\label{eq:bms1},
\end{equation}
and are explicitly given by 
\begin{equation}
\xi^u=f,\quad 
\xi^\phi  =  Y - \partial_\phi f \,\int_r^\infty dr' \, {r'}^{-2}
e^{2\beta},\quad 
\xi^r =- r(\partial_\phi \xi^\phi-U\partial_{\phi} f), \label{eq:ads3vect}
\end{equation}
where $\partial_r f =\partial_r Y = 0$. 
When requiring in addition that the fall-off conditions
\eqref{eq:fall1} be preserved, 
\begin{equation}
\mathcal{L}_\xi g_{ur}=O(r^{-1}),\quad \mathcal{L}_\xi g_{u\phi}=O(1), \quad
\mathcal{L}_\xi g_{uu}=O(1).\label{eq:bms2}
\end{equation}
the additional conditions are 
\begin{equation}
\partial_u f=\partial_\phi Y,\quad \partial_u
Y=\frac{1}{l^2}\partial_\phi f, \label{eq:120}
\end{equation}
and hence
\begin{equation}
f=\frac{l}{2}(Y^++Y^-), \quad
Y=\frac{1}{2}(Y^+-Y^-), \label{eq:Y}
\end{equation}
where now $x^\pm=\frac{u}{l}\pm\phi$, and $Y^\pm=Y^\pm (x^\pm)$ are
arbitrary functions of their arguments. It now follows that
\eqref{eq:44a}, or equivalently \eqref{eq:1}, also hold in the BMS
gauge: the spacetime vectors \eqref{eq:ads3vect} form a representation
of the conformal Lie algebra on an asymptotically $AdS_3$ spacetime
defined through (\ref{eq:ojo}) and \eqref{eq:fall1} when equipped with
the modified bracket. 

In the gauge \eqref{eq:ojo} assuming furthermore that $\beta=o(1)=U$,
the most general solution to the Einstein equations $G_{\alpha
  \beta}=l^{-2}g_{\alpha \beta}$ is easily worked out. One finds 
\begin{equation}
  \label{eq:10}
  \beta=0,\quad U=-r^{-2} \mathcal{N},\quad
  \frac{V}{r}=-\frac{r^2}{l^2}+\mathcal{M}-r^{-2} \mathcal{N}^2,
\end{equation}
or equivalently
\begin{equation}
ds^2=\left(-\frac{r^2}{l^2}+\mathcal{M}\right)du^2-2du
  dr+2\mathcal{N}du d\phi+r^2d\phi^2\label{eq:Squad},
\end{equation}
where $\d_r \mathcal{M}=0=\d_r \mathcal{N}$. In addition, 
\begin{equation}
\partial_u \mathcal{M}
=\frac{2}{l^2}\partial_{\phi}\mathcal{N}, \quad 2\partial_u \mathcal{N} = \partial_{\phi}
\mathcal{M},\label{eq:13}
\end{equation}
so that 
\begin{equation}
\mathcal{M}(u,\phi)=2(\Xi_{++}+\Xi_{--}),\quad 
\mathcal{N}(u,\phi)=l(\Xi_{++}-\Xi_{--}),\label{MN}
\end{equation}
where $\Xi_{\pm\pm}=\Xi_{\pm\pm}(x^\pm)$. Note that on-shell, the
asymptotic Killing vectors become 
\begin{equation}
\xi^u=f,\quad \xi^\phi=Y-r^{-1}\d_\phi f,\quad \xi^r=-r\d_\phi
Y+\d^2_\phi f-r^{-1} \mathcal{N} \d_\phi f.\label{eq:18}
\end{equation}

In these coordinates, the BTZ black hole is again determined by
\eqref{BTZ}, while computing $\mathcal{L}_\xi g_{\mu\nu}$, with $\xi$
as in \eqref{eq:18} and $g_{\mu\nu}$ as in \eqref{eq:Squad} and
\eqref{MN}, gives the transformation laws \eqref{eq:A15}.

The charges are computed along the lines of \cite{Barnich:2001jy} at
the circle at infinity $u$ constant and $r \rightarrow \infty$ using as
background the $AdS_3$ space-time in the form
 \begin{equation}
  \label{eq:qw}
  d\bar s^2=-\left(1+\frac{r^2}{l^2}\right)du^2-2dudr+r^2 d\phi^2.
\end{equation}
In a first stage, this gives 
\begin{equation}
  Q_{\xi}[h,\bar{g}]=\frac{1}{16 \pi G} \int_0^{2\pi}d\phi\, \big[f (\mathcal{M}+1) +
  2Y \mathcal{N}\big],  \label{eq:qq} 
\end{equation}
and then, using \eqref{eq:Y} and \eqref{MN}, one recovers
\eqref{eq:2}. It follows that all relations of the previous section
discussed after equation \eqref{eq:2} hold in the BMS gauge as well.

\section{Flat Limit}\label{lim}

We now study the flat limit of asymptotically $AdS_3$ space-times in
the BMS gauge. More precisely, we want to take the limit of the
cosmological constant going to zero in such a way that asymptotically
$AdS_3$ solutions get mapped to asymptotically flat solutions. This
will be achieved through a suitably modified Penrose limit.

The first observation is that asymptotically $AdS_3$ spacetimes in the
BMS gauge belong to the Robinson-Trautman class of metrics
\cite{RobinsonTrautman2}, for which $u$ parametrizes null
hypersurfaces and $r$ is an affine parameter for the associated null
geodesic generators. It is in this setting that Penrose has shown that
a particular limit of a generic spacetime is given by a plane wave
\cite{Penrose:1976xx}. Second, in the context of string theory backgrounds,
Penrose limits were used to connect backgrounds with $AdS$ factors to
backgrounds involving flat or plane wave backgrounds
\cite{0264-9381-19-10-101,Blau:2002mw}. 

Consider the Einstein-Hilbert action for three dimensional anti-de
Sitter gravity,
\begin{equation}
  \label{eq:22}
  S[g;G,l]=\frac{1}{16\pi G}\int d^3x\sqrt{|g|}( R+\frac{2}{l^2}). 
\end{equation}
For any metric $g$ and $\lambda>0$, let
$g^{(\lambda)}=\lambda^{-2}g$ denote the suitably rescaled metric.
If $G^{(\lambda)}=\lambda^{-1}G$, $l^{(\lambda)}=\lambda^{-1}l$, we have 
\begin{equation}
  \label{eq:23}
  S[g^{(\lambda)};G^{(\lambda)},l^{(\lambda)}]=S[g;G,l].
\end{equation}

Consider then a family of spacetimes $g(\lambda)$ labelled by a
parameter $\lambda>0$ that is a solution to the Einstein equations
with cosmological constant $\Lambda=-\frac{1}{l^2}$, $R_{\mu
  \nu}[g(\lambda)]=-2l^{-2}g_{\mu \nu}(\lambda)$. In particular,
according to the previous section, the most general asymptotically
$AdS_3$ family of solutions of this type is obtained by using
$\Xi_{\pm\pm}(\lambda)=\Xi_{\pm\pm}(x^\pm,\lambda)$ with an arbitrary
$\lambda$ dependence instead of $\Xi_{\pm\pm}(x^\pm)$ in \eqref{MN}
and \eqref{eq:Squad}.  It then follows from \eqref{eq:23} that
$g^{(\lambda)}(\lambda)$ is a solution to the Einstein equations with
cosmological radius $l^{(\lambda)}$. This can also be seen directly by
using 
\begin{eqnarray}
\label{scaling}
R_{\mu \nu}[\lambda^{-2}g(\lambda)]=R_{\mu \nu}[g(\lambda)]=
-2\frac{\lambda^2}{l^2}(\lambda^{-2}g_{\mu \nu}(\lambda)).
\end{eqnarray}
If $g= \lim_{\lambda \rightarrow 0} \lambda^{-2} g(\lambda)$ is a
well-defined metric, it thus defines a solution to the Einstein
equations with vanishing cosmological constant.

For instance, the Penrose limit \cite{Penrose:1976xx} of the metrics
\eqref{eq:Squad} consists in rescaling the metric as above and
simultaneously scaling the coordinates as
\begin{eqnarray}
\label{Penrosescaling}
(u,r,\phi) \rightarrow (\lambda^2 u, r, \lambda \phi).
\end{eqnarray}  
The transformed metric is
\begin{eqnarray}
\label{scg}
\lambda^{-2} ds^2_{\lambda}=-\lambda^2\Big[\frac{r^2}{l^2}-
\mathcal{M}(\lambda^2 u,\lambda \phi)\Big]du^2-2dudr+
2 \lambda \mathcal{N}(\lambda^2u, \lambda\phi) du d\phi + r^2 d{\phi}^2.
\end{eqnarray}
If $\mathcal{M}(u,\phi)$ and $\mathcal{N}(u,\phi)$ are continuous functions of the
coordinates, the Ricci-flat limiting metric is simply the null
orbifold \cite{Horowitz:1990ap}
\begin{eqnarray}
\label{scga}
d{s}^2=-2dudr + r^2 d{\phi}^2.
\end{eqnarray}  

We would like to perform a different flat limit\footnote{It should be
  interesting to analyze such modified Penrose limits in higher
  dimensions.} which keeps the number of arbitrary functions
appearing in the metric. This can be done through the coordinate
scaling,
\begin{eqnarray}
\label{Newscaling}
(u,r,\phi) \rightarrow (\lambda u, \lambda r, \phi).
\end{eqnarray}  
In this case, the transformed metric is
\begin{eqnarray}
\label{scg2}
\lambda^{-2} ds^2_{\lambda}=\Big[-\frac{\lambda^2 r^2}{l^2}
+\mathcal{M}(\lambda u,\phi)\Big]du^2-2dudr+
2\lambda^{-1}\mathcal{N}(\lambda u, \phi) du d\phi + r^2 d{\phi}^2.
\end{eqnarray} 
In order to control what happens in the limit, we have to take into
account the expression \eqref{MN} of the functions
$\mathcal{M},\mathcal{N}$ in terms of the arbitrary functions
$\Xi_{\pm\pm}(\lambda)$. In terms of modes, this gives
\begin{equation}
\begin{gathered}
  \label{eq:14}
  \mathcal{M}(\lambda u,\phi)=
-1+8G \sum_m\left(\frac{L^+_{m}(\lambda) e^{-im\frac{\lambda
          u}{l}} + L^-_{-m}(\lambda) e^{im\frac{\lambda
          u}{l}}}{l}\right) e^{-im\phi},\\
\lambda^{-1} \mathcal{N}(\lambda u,\phi)=
\frac{4G}{\lambda}\sum_m\left( L^+_{m}(\lambda)
  e^{-im\frac{\lambda u}{l}} - L^-_{-m}(\lambda) e^{im\frac{\lambda
      u}{l}}\right) e^{-im\phi}. 
\end{gathered}
\end{equation}
In other words, up to the arbitrary $\lambda$ dependence in
$\Xi_{\pm\pm}$, or equivalently in their Fourier modes, in the
coordinates $u,r,\phi$, the metric 
$\lambda^{-2}ds^2_\lambda$ is obtained from \eqref{eq:Squad} simply by
replacing $l\to l^{(\lambda)}$ and $G\to G^{(\lambda)}$. 

In order to have a well-defined limit, we then need that
\begin{equation}
\begin{gathered}
L^+_m(\lambda)=\half
 lP_m+\lambda L^{\prime +}_m(0) +O(\lambda^2),\\\label{eq:15}
L^-_m(\lambda)=\half
lP_{-m}+\lambda L^{\prime -}_m(0) +O(\lambda^2).
\end{gathered}
\end{equation}
In this case, the limit becomes  
\begin{equation}
\begin{gathered}
  \label{eq:16}
   \lim_{\lambda\to 0}\mathcal{M}(\lambda u,\phi)=-1+8G \sum_m P_{m}
   e^{-im\phi}=\Theta(\phi), \\
\lim_{\lambda\to 0}\lambda^{-1} \mathcal{N}(\lambda u,\phi)=4G \sum_m\left(J_{m}
-u im P_{m}\right)e^{-im\phi}=\Xi(\phi)+\frac{u}{2}\partial_\phi\Theta(\phi),  
\end{gathered}
\end{equation}
with 
\begin{equation}
P_m=\frac{L^+_m(0)+L^-_{-m}(0)}{l},\quad
J_m=L^{\prime +}_m(0)-L^{\prime -}_{-m}(0).\label{eq:20}
\end{equation}
In summary 
\begin{equation}
  \label{eq:12}
  \lim_{\lambda\to
    0}\lambda^{-2}ds^2_\lambda=\Theta(\phi)du^2-2dudr+
2\Big[\Xi(\phi)+\frac{u}{2}\partial_{\phi}\Theta(\phi)\Big]du
  d\phi + r^2 d{\phi}^2.  
\end{equation}
We have thus recovered through this limit the most general solution to
the asymptotically flat Einstein equations as discussed in
\cite{Barnich:2010eb}, i.e, Ricci flat metrics metrics of the form
\eqref{eq:ojo} with the fall-off conditions as in \eqref{eq:fall1}
with $l\to\infty$ so that $\frac{V}{r}=O(1)$.

The limit can be used, as well, to relate the symmetries of
asymptotically $AdS_3$ and flat spacetimes. The first step is again
that for a general asymptotically $AdS_3$ metric $g(\lambda)$, the
most general asymptotic Killing vectors involve the arbitrary
functions $Y^{\pm}(x^\pm,\lambda)=\sum_m
Y^\pm_{-m}(\lambda)e^{imx^\pm}$ with an arbitrary dependence on $\lambda$
in \eqref{eq:FGvect} instead of $Y^\pm(x^\pm)$. Under the change of
coordinates \eqref{Newscaling}, the asymptotic Killing vectors
\eqref{eq:18} 
then acquire the form
\begin{equation}
\xi{(\lambda)}=\lambda^{-1} \xi^{u}(\lambda u, 
\lambda r, \phi)\partial_{u}+\lambda^{-1}  
\xi^{r}(\lambda u, \lambda r, \phi)\partial_{r}+ 
\xi^{\phi}(\lambda u, \lambda r, \phi)\partial_{\phi},
\end{equation} 
 Consider first the leading order parts of $\xi^{u}(\lambda)$ and
 $\xi^{\phi}(\lambda) $, 
\begin{equation}
  \label{eq:17}
  \begin{gathered}
\lambda^{-1}f(\lambda u,\phi)=\frac{l}{2\lambda}\sum_m
\left(Y^+_{m}(\lambda) e^{-im\frac{\lambda u}{l}} +Y^-_{-m}(\lambda)
  e^{im\frac{\lambda u}{l}}\right) e^{-im\phi},\\
Y(\lambda u,\phi)=\half \sum_m
\left(Y^+_{m}(\lambda) e^{-im\frac{\lambda u}{l}} -Y^-_{-m}(\lambda)
  e^{im\frac{\lambda u}{l}}\right) e^{-im\phi}.
  \end{gathered}
\end{equation}
When taking into account the scaling of $r$ and the previously
discussed behaviour of $\mathcal{N}$ in \eqref{eq:14}, it follows again that,
apart from the arbitrary $\lambda$ dependence in the Fourier modes, in
the coordinates $u,r,\phi$, $\xi(\lambda)$ is obtained from
\eqref{eq:18} through the substitution $l\to l^{(\lambda)}$, $G\to
G^{(\lambda)}$.

A necessary condition for a well defined limit 
$\lambda\to 0$ is then 
\begin{equation}
\label{Texp}
\begin{gathered}
 Y_{m}^{+}(\lambda)=Y_{m}+\lambda Y^{\prime +}_{m}(0) 
+ O(\lambda^2),\\
 Y_{m}^{-}(\lambda)=-Y_{-m}+ \lambda Y^{\prime -} _{m}(0)
+ O(\lambda^2), 
\end{gathered}
\end{equation}
so that 
\begin{equation}
\label{Texp+}
\begin{gathered}
\lim_{\lambda\to 0}\lambda^{-1}f(\lambda u,\phi)=\sum_m \left(T_{m}
-uim Y_{m}\right) e^{-im\phi}=T(\phi)+u\d_\phi Y(\phi),\\
\lim_{\lambda\to 0} Y(\lambda u,\phi)=\sum_m
Y_{m} e^{-im\phi}=Y(\phi),
\end{gathered}
\end{equation}
with 
\begin{equation}
T_m=\frac{l}{2}\left(Y^{\prime +}_m(0)+Y^{\prime
    -}_{-m}(0)\right)\label{eq:21},\quad 
Y_m=\half\left (Y^+_m(0)-Y^-_{-m}(0)\right).
\end{equation}
Again, the scaling of $r$ and the previously discussed limit of $\mathcal{N}$ in
\eqref{eq:16}, then implies that $\lim_{\lambda\to 0}
\xi{(\lambda)}=\xi^\mu_{T,Y}\d_\mu$ where the components
$\xi^\mu_{T,Y}$ are given by \eqref{eq:18} with $f=T+u\partial_\phi Y$
and $\mathcal{N}=\Xi+\frac{u}{2}\d_\phi\Theta$. This limit coincides with
the direct computation in \cite{Barnich:2010eb} of the vectors
describing the symmetries of asymptotically flat spacetimes in three
dimension. 

Let us now turn to the charges. In the presence of $\lambda$, the
charges $Q_{Y^+,Y^-}(\lambda)$ are given by \eqref{eq:qq} with
$f(u,\phi)\to \lambda^{-1} f(\lambda u,\phi)$, $Y(u,\phi)\to Y(\lambda
u,\phi)$, $\mathcal{M}(u,\phi)\to \mathcal{M}(\lambda u,\phi)$ and 
$\mathcal{N}(u,\phi)\to
\lambda^{-1}\mathcal{N}(\lambda u,\phi)$. In the limit, one finds
\begin{equation}
  \label{eq:11b}
  \lim_{\lambda\to 0} Q_{Y^+,Y^-}{(\lambda)}=\frac{1}{16 \pi G}
  \int_0^{2\pi}d\phi\, \big[T (\Theta+1) +
  2Y \Xi\big]=Q_{T,Y}\,
\end{equation}
in agreement with the direct computation in
\cite{Barnich:2010eb}\footnote{Note that in the last line of (3.18) of
  this reference $\Theta$ should be replaced by $\Theta+1$.}.
In terms of modes, we have 
\begin{equation}
  \label{eq:19}
  Q_{Y^+,Y^-}{(\lambda)}=\lambda^{-1}\sum_m \left(
    Y^+_{-m}(\lambda)L^+_m(\lambda)+Y^-_{-m}(\lambda)L^-_m(\lambda)\right)
\stackrel{\lambda\to
    0}{\longrightarrow}\sum_{m} \left(T_{-m}P_m+Y_{-m}J_{m}\right), 
\end{equation}
by taking into account \eqref{eq:20} and \eqref{eq:21}.

In the limit, the algebra of the vectors $\xi^\mu_{T,Y}\d_\mu$ in the
Lie algebroid bracket $[\cdot,\cdot]_M$ has been computed directly and
shown to form a representation of $\mathfrak{bms}_3$. More concretely,
in terms of this bracket, the vectors
$t_m=\xi^\mu_{e^{im\phi},0}\d_\mu$, $l_m=\xi^\mu_{0,e^{im\phi}}\d_\mu$
satisfy \eqref{eq:qflatbis} with $c_1=0=c_2$. Similarly, the Dirac
bracket algebra has also been computed directly in the limit. If
$P_m=Q_{e^{im\phi},0}$ and $J_m=Q_{0,e^{im\phi}}$, their Dirac bracket
algebra is \eqref{eq:qflatbis} with central charges given in
\eqref{eq:8}. It thus follows that the asymptotic symmetry algebra and
its representation get contracted in the limiting procedure, as also
observed for exact isometries in \cite{Blau:2002mw} for instance.

For completeness, let us finish by showing how the contraction of the
algebra occurs concretely in this case. For definiteness, we will
concentrate on the vector fields 
\begin{equation}
  \label{eq:11}
 v_{Y^+,Y^-}= f\partial_u+Y\partial_\phi, 
\end{equation}
with $f,Y$ given in terms of $Y^+,Y^-$ as in \eqref{eq:Y}. These
vectors fields form a representation of (two commuting copies) of the
conformal algebra in the standard Lie bracket. As before, besides the
arbitrary $\lambda$ dependence, in the coordinates $u,\phi$,
$v_{Y^+,Y^-}(\lambda)$ is obtained by replacing
$l\to l^{(\lambda)}$. It then easily follows that
$v_{Y^+,Y^-}(\lambda)$ also form a representation of the conformal
algebra,
\begin{equation}
  \label{eq:24}
  [v_{Y^+_1,Y^-_1}(\lambda),v_{Y^+_2,Y^-_2}(\lambda)]=v_{[Y_1,Y_2]^+,[Y_1,Y_2]^-}(\lambda).
\end{equation}
This can be seen in the coordinates
$x^{\pm}(\lambda)=\frac{u\lambda}{l}\pm \phi$, where
$v_{Y^+,Y^-}(\lambda)=Y^+(\lambda,x^+(\lambda))\d_{x^+(\lambda)}+
Y^-(\lambda,x^+(\lambda))\d_{x^-(\lambda)}$. In particular, on the
right hand side of \eqref{eq:24}, we have $[Y_1,Y_2]^\pm(\lambda)
=Y_1^\pm(\lambda,x^\pm(\lambda))\d_{x^\pm(\lambda)}
Y^\pm_2((\lambda,x^\pm(\lambda))-(1\leftrightarrow 2)$. We have
already shown that, when the expansion \eqref{Texp} holds,
$\lim_{\lambda\to 0} v_{Y_1,Y_2}(\lambda)=v_{T,Y}$ with $v_{T,Y}$ of
the form \eqref{eq:11} with $f=T+u\d_\phi Y$ and $T,Y$ arbitrary
functions of $\phi$. We then need to evaluate the right hand
side. Before taking the limit, the $u$ and $\phi$ components are given
by
\begin{equation}
  \label{eq:25}
\begin{gathered}
  \frac{l}{2\lambda}\sum_{n,m} i(m-n) \left[ Y^+_{m 1}(\lambda)Y^+_{n
      2}(\lambda) e^{-i(m+n)\frac{u\lambda}{l}}-Y^-_{-m 1}(\lambda)Y^-_{-n
      2}(\lambda)
    e^{i(m+n)\frac{u\lambda}{l}}\right]e^{-i(m+n)\phi},\\
\frac{1}{2}\sum_{n,m} i(m-n) \left[ Y^+_{m 1}(\lambda)Y^+_{n
      2}(\lambda) e^{-i(m+n)\frac{u\lambda}{l}}+Y^-_{-m 1}(\lambda)Y^-_{-n
      2}(\lambda)
    e^{i(m+n)\frac{u\lambda}{l}}\right]e^{-i(m+n)\phi}.
\end{gathered}
\end{equation}
Using \eqref{Texp}, taking the limit and reconstructing the functions
gives $Y_1\d_\phi T_2+T_1\d_\phi Y_2+u \d_\phi (Y_1\d_\phi Y_2)
-(1\leftrightarrow 2)$ and $Y_1\d_\phi Y_2-(1\leftrightarrow 2)$, as
one should for the $\mathfrak{bms}_3$ algebra.

\section{Zero modes}

Let us now concentrate on the zero modes. In the AdS case, from
\eqref{eq:Squad}, \eqref{MN}, in the parametrization in terms of $M$
and $J$ as in \eqref{BTZ}, we
thus have $\mathcal{M}=8G M$, $\mathcal{N} =4GJ$ with charges
\begin{equation}
Q_{\partial_u}=M+\frac{1}{8G},\quad Q_{\partial_\phi}=J\label{eq:5}.
\end{equation}
This metric is explicitly related to the standard ADM form,
\begin{equation}
  \label{eq:26}
\begin{gathered}
  ds^2=-N^2dt^2+N^{-2}dr^2+r^2(d\varphi+N^\varphi dt)^2,\\ 
N^2=\frac{r^2}{l^2}-8MG +\frac{16 G^2 J^2}{r^2},\quad
N^\varphi=\frac{4GJ}{r^2},
\end{gathered}
\end{equation}
through $t=u+f(r)$, $\varphi=\phi+g(r)$, where $f^\prime= N^{-2}$,
$g^\prime=-N^\varphi f^\prime$. In turn, the Fefferman-Graham form
\eqref{FG} for constant $\Xi_{\pm\pm}$ is then obtained through
$x^\pm=\frac{t}{l}\pm\varphi$ and $\frac{r^2}{l^2}=\frac{\rho^2}{l^2}
+(\Xi_{++}+\Xi_{--}) +\frac{l^2}{\rho^2}\Xi_{++}\Xi_{--}$. 

Let us first comment briefly on the space of solutions with
non-vanishing negative cosmological constant, depicted in Fig. 2a. The
BTZ black holes correspond to the region $M\geq
\frac{|J|}{l}$. Geometries satisfying $0<M< \frac{|J|}{l}$ leave
exposed the chronological singularity at $r=0$ which encloses an
unbounded region, $r < 0$, containing closed time-like
curves\cite{Banados:1993gq}.  On the other hand, for
$-\frac{1}{8G}<M\leq-\frac{|J|}{l}$, the geometry describes a spinning
particle sitting at $r=0$ \cite{Deser:1983dr}, which produces a conical defect
around it (see also \cite{Miskovic:2009uz}).  In the neighborhood of
the particle, there is, in general, a bounded region containing
closed timelike curves.  When $M=-\frac{1}{8G}$, the angular defect
vanishes and the geometry is smooth everywhere.  The case $J=0$
describes global AdS spacetime.  Below this mass, the angular defect
becomes an excess.  When $0>M>-\frac{|J|}{l},$ the whole spacetime
contains closed time-like curves \cite{Brown:1986nw}.

In the flat case, we have the same metric \eqref{eq:26} without the
$\frac{r^2}{l^2}$ term, or, in null coordinates,  the right hand side of \eqref{eq:12}
with $\Theta=8GM$ and $\Xi=4GJ$, so that the charges are the same as
in the AdS case, and thus again given by \eqref{eq:5}.

Identifying in both cases the solutions with the same charges then
implies that $AdS_3$ corresponds to the Minkowski space-time
$M=-\frac{1}{8G}$, $J=0$, while $M=0=J$, the massless BTZ black 
hole, corresponds to the null orbifold \eqref{scga}. For $8GM=-\alpha^2<0$,
we can supplement the change of variable that leads to
\eqref{eq:26} by the redefinition of the radial coordinate $\bar{r}^2
=\frac{r^2}{\alpha^2}+\frac{16G^2J^2}{\alpha^4}$, bringing the line 
element into
\begin{eqnarray}
\label{spinningparticle}
ds^2= -\left(\alpha dt-\frac{4GJ}{\alpha}d\varphi \right)^2+d\bar{r}^2+
\alpha^2 \bar{r}^2 d{\varphi}^2,
\end{eqnarray} 
which corresponds to a spinning particle in flat spacetime
\cite{Deser:1983tn}.  Under the redefinition $\bar{t}=\alpha
t-\frac{4GJ}{\alpha}\varphi$ and $\bar{\phi}=\alpha \varphi$, the
identification $(t,\varphi)\sim(t,\varphi+2\pi)$ becomes
$(\bar{t},\bar{\phi})\sim(\bar{t}-\frac{8\pi G
  J}{\alpha},\bar{\phi}+2\pi\alpha )$.  For $\alpha^2<1$, the spatial
geometry corresponds to a cone with deficit angle $\Delta
\bar{\phi}=2\pi (1-|\alpha|)$, while for $\alpha^2=1$ and $J=0$, the
geometry corresponds to global Minkowski spacetime. For
$\alpha^2>1$, the geometries possess an angular excess. This family
may have arbitrarily negative values of $M$.
 
We now pass to study the case of positive $M$, and define
$8GM=\alpha^2$.  It is convenient to separate the analysis for
$r<\frac{4G|J|}{|\alpha|}$ and $r>\frac{4G|J|}{|\alpha|}$. In the
first region, we make the transformation
$\bar{r}^2=-\frac{r^2}{\alpha^2}+\frac{16G^2J^2}{\alpha^4}$ which
produce the line element,
\begin{eqnarray}
\label{CTCspacetime}
ds^2= \left(-\alpha dt+\frac{4GJ}{\alpha}d\varphi \right)^2+d\bar{r}^2
-\alpha^2 \bar{r}^2 d{\varphi}^2.
\end{eqnarray} 
Therefore, inside this region,  the direction $\partial_{\varphi}$ is always
time-like, generating closed time-like curves. It is a bounded time machine.
For $r>\frac{4G|J|}{\alpha}$, the suitable transformation is
$\bar{r}^2=\frac{r^2} {\alpha^2}-\frac{16G^2J^2}{\alpha^4}$, and the
metric turns out to be
\begin{eqnarray}
\label{cosmology}
ds^2= -dT^2+\left(\frac{4GJ}{\alpha}\right)^2 dX^2
+\alpha^2 T^2 d{\varphi}^2,
\end{eqnarray}
where we have defined $T=\bar{r}$, $X=\frac{\alpha^2}{4GJ} t + \varphi
$ and and we have made the identification $X \sim X+2\pi$ and $\varphi
\sim \varphi+2\pi$.  The outcoming spacetime is a cosmology whose
spatial section is a 2-torus with radii $\frac{4GJ}{\alpha}$ and
$\alpha T$. Note that here the parameter $M$ may not be identified
with a mass. It is conjugate to a space-like translation generator,
and corresponds to a momentum. When $J=0$, the metric becomes $ds^2=
-dT^2+ dX^2 +\alpha^2 T^2 d{\varphi}^2$ with an unwrapped
$X$-coordinate.

\begin{figure}[h]
\label{fig2} 
\begin{center}
  \includegraphics[width=15cm]{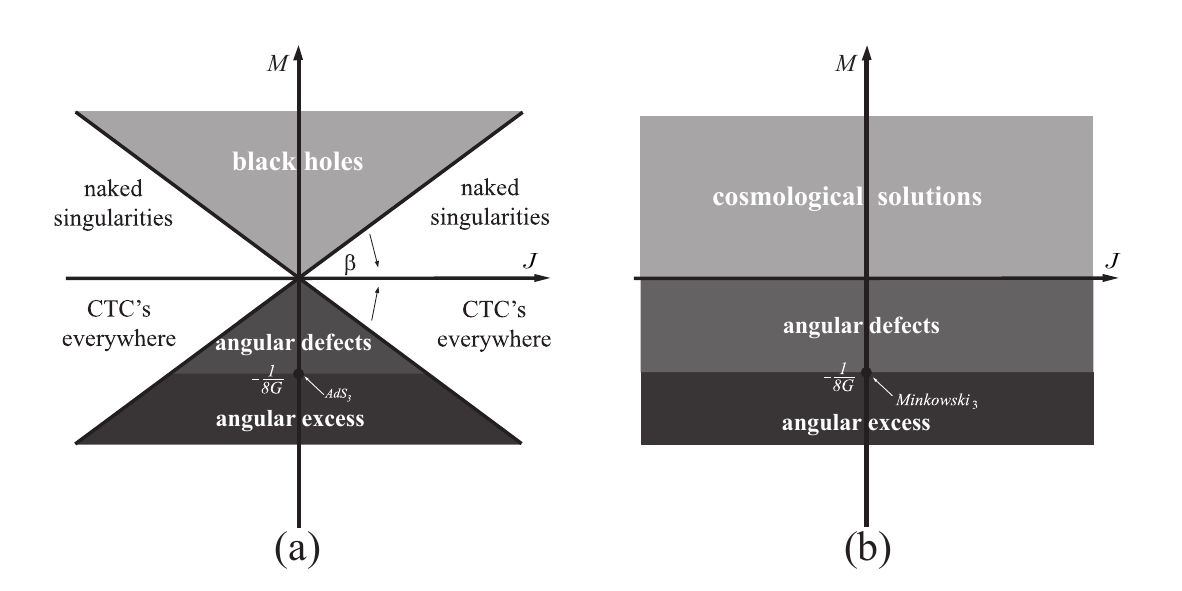}
  \caption{Zero mode solutions of 2+1 gravity. Figure (a) depicts the
    case of non-vanishing cosmological constant. The slope $\beta$ is
    given by $\tan \beta=\frac{1}{l}$. Figure (b) shows the limit $l
    \to \infty$, when $\beta$ vanishes. No solutions are lost in the
    limit, but the horizon of the BTZ black holes gets pushed to
    infinity, hence the time coordinate becomes spatial everywhere and
    the line element describes the non-static, cosmological solution
    (\ref{cosmology}). }
   \end{center}
\end{figure}

\section*{Acknowledgements}
\label{sec:acknowledgements}

\addcontentsline{toc}{section}{Acknowledgments}

The work of G.B.~is supported in part by the Fund for Scientific
Research-FNRS (Belgium), by the Belgian Federal Science Policy Office
through the Interuniversity Attraction Pole P6/11, by IISN-Belgium, by
``Communaut\'e fran\c caise de Belgique - Actions de Recherche
Concert\'ees'' and by Fondecyt Projects No.~1085322 and
No.~1090753. The work of AG was partially supported by Fondecyt
(Chile) Grant \#1090753.

\def\cprime{$'$}
%

\end{document}